\documentclass[12pt]{iopart}
\usepackage{psfig,graphicx,epsf,tabularx}

\begin{document}
\title{Influence of Co$^{3+}$ spin-state on optical properties of LaCoO$_3$ and HoCoO$_3$}
\author{L.V.~Nomerovannaya$\dag$,  A.A.~Makhnev$\dag$, S.V.~Streltsov$\dag\ddag$,
I.A.~Nekrasov$\dag$, M.A.~Korotin$\dag$,  S.V.~Shiryaev$\S$, G.L.~Bychkov$\S$,
S.N.~Barilo$\S$ and V.I.~Anisimov$\dag$}
\address{$\dag$Institute of Metal Physics, 620219 Ekaterinburg GSP-170, Russia}
\address{$\ddag$Ural State Technical University, 620002 Ekaterinburg, Russia}
\address{$\S$Institute of Solid State and Semiconductor Physics, 220072 Minsk. Belarus}

\ead{Streltsov@optics.imp.uran.ru}

\begin{abstract}
Optical properties of the isoelectronic compounds LaCoO$_3$ and
HoCoO$_3$ has been experimentally and theoretically investigated.
We've measured the real $\varepsilon_1(\omega)$ and imaginary
$\varepsilon_2(\omega)$ parts of the dielectric function,
reflectance $R(\omega)$ and optical conductivity at room
temperature. The shift of the most pronounced spectral features to
the high energy region on 0.3 eV associated with larger
distortions due to the smaller rare earth ionic radii in HoCoO$_3$
in comparison with LaCoO$_3$ was observed. Also there was found
an enhancement of absorption intensity in the range 1.3-2.3 eV
in all kinds of spectra in HoCoO$_3$, which can be attributed basing on 
the results of LDA+U calculations to
the different spin-states of Co$^{3+}$ ion in these compounds.
The shift of the onset of the absorption from less than 0.1 eV in 
LaCoO$_3$ to 0.7 eV in
HoCoO$_3$ and an absorption intensity enhancement in a narrow
spectral range 1.2-2.6 eV in HoCoO$_3$ are clearly seen from the 
calculated convolution of partial densities of states obtained in
the LDA+U approach. Such changes are assumed to be induced
by the different Co$^{3+}$ spin-state in these compounds at room
temperature.
\end{abstract}

\submitto{\JPCM}
\maketitle

\section{Introduction}

At present a temperature dependence of
structural\cite{Radaelli-02,Maris} and
magnetic\cite{Asai-98,Zobel-02,Kobayashi-00} properties of
LaCoO$_3$ is usually described within a three-spin-states model.
In the ground state LaCoO$_3$ is a nonmagnetic insulator and all
Co$^{3+}$ ions have a low spin-state configuration (LS,
t$_{2g}^6$e$_g^0$, S=0)\cite{Imada-98}.

There are two transitions at approximately 100 and 500~K in
LaCoO$_3$ with the increase of temperature. At 100~K this compound
undergoes the spin-state transition to a paramagnetic state, as
evidenced by a steep increase of magnetic
susceptibility\cite{Imada-98} and can be interpreted as a
spin-state transition to the intermediate spin-state (IS,
t$_{2g}^5$e$_g^1$, S=1) or to a mixture of low and high spin-state
(HS, t$_{2g}^4$e$_g^2$, S=2)\cite{Goodenough-58}. The second
transition at 500~K is associated with a metal-insulator (MI)
transition. Band structure calculations within LDA+U approximation
demonstrated that IS is the lowest in energy spin-state after the
first transition\cite{Korotin-96,Ravindran-02,Nekrasov-03}. 
In contrast to the expectation from the simple
ionic model IS is stabilized by a strong $p-d$ hybridization and
possible orbital ordering in $e_g$ shell of Co$^{3+}$ ions
\cite{Korotin-96}.

From experimental point of view there are gradual changes in x-ray
absorption spectroscopy\cite{Saitoh-97,Hu-03} at the first
spin-state transition. Up to now optical spectroscopy as well as
an electroresistivity investigation have not found any significant
changes in the electronic structure of LaCoO$_3$ in the low
temperature region\cite{Tokura-98}. Change of the optical
conductivity spectrum was observed only near MI transition.
However, infrared spectroscopy revealed anomalous splitting of the
phonon modes and change of their intensities with the increase of
temperature, which was associated with local distortions due to
the spin-state transitions\cite{Yamaguchi-97}.

In this paper we've performed experimental and theoretical investigations
of the changes in electronic structure, occurring at the spin-state
transition. We've chosen two isoelectronic compounds: LaCoO$_3$
and HoCoO$_3$, which have different spin-state configurations at 
room temperature due to a chemical pressure happening with substitution of La
ions by smaller Ho ions and hence increasing of a crystal field
$t_{2g}-e_{g}$ energy splitting \cite{Nekrasov-03}.
Observed changes in the optical properties of
HoCoO$_3$ in comparison with LaCoO$_3$ can be associated
with larger lattice distortions in HoCoO$_3$ and with different spin-state
stabilization of Co$^{3+}$ ions in these compounds.

\section{Experimantal details}

In order to grow LaCoO$_3$ and HoCoO$_3$ single crystals a kind of anodic electrodeposition technique was used.
In particular, McCarrol approach \cite{McCarrol-97} was modified \cite{Shiryaev-01}
to use seeded flux melt growth based on Cs$_2$MoO$_4$-MoO$_3$ mixture in the ratio 2.2 : 1 as solvent.
Appropriate solute quantity was added into a 100 $cm^3$ platinum crucible contained the mixture to
grow these single crystals with a seed served as anode at
$\sim$ 950-1000 $^{\circ}$C under current density in the range
0.5-0.7 mA/cm$^2$. Simultaneously the crucible serves as cathode in a such electrochemical cell.

The typical dimensions of the samples were 2x2x1 mm$^3$.
According to X-ray data they were single phase.
Optical measurements were
performed on a cleaved mirror as well as on mechanically
polished by diamond powder (with grains less than 0.5 $\mu$m) surface.
In this paper we use experimental data for a cleaved surface, because
in this case numerical data for the reflectance were higher and
spectral features had a better resolution.

The refractive index $n$ and absorption constant $k$ were measured in the
spectral range 0.5-5.0 eV by ellipsometric technique at the
room temperature. The automated ellipsometer for measurements on
small samples was assembled on the basis of the KSVU-12 spectrometer.
For energies less than 0.5 eV absorption constant $k$ abruptly decreases
and an error of measured values rises. Thus we had to break off
our measurements at 0.5 eV. The relative experimental errors were 2$\%$-4$\%$.
From $n$ and $k$, the real $\varepsilon_1(\omega)=n^2-k^2$ and imaginary
$\varepsilon_2(\omega)=2nk$ parts of the complex dielectric constant
$\varepsilon(\omega)$, the optical conductivity
$\sigma(\omega)=nk\omega/2\pi$ and the reflectance
$R(\omega)=[(n-1)^2+k^2]/[(n+1)^2+k^2]$ were derived.

\section{Experimantal results}

The real $\varepsilon_1(\omega)$ and  imaginary $\varepsilon_2(\omega)$
parts of the dielectric function for LaCoO$_3$ and HoCoO$_3$ are presented
in figure 1. As seen from behaviour of $\varepsilon_2(\omega)$ function, there is
a strong absorption region in the energy range 1.0-5.0 eV for
both compounds. Two maxima for HoCoO$_3$ are clearly observed
at 1.7 and 2.9 eV. For LaCoO$_3$ such features are broaden and shifted
to the low energy region on $\sim$ 0.3 eV. The dispersion of $\varepsilon_1(\omega)$
near these peaks has anomalous behavior and positive values show that optical
properties are defined by interband transitions. Another one
feature is revealed more distinctly in the spectrum of the reflectance
R$(\omega)$ for HoCoO$_3$, which has a maximum at 4.5 eV (figure 1, inset).

\begin{figure}
\begin{center}
 \epsfysize=110mm
 \epsfbox{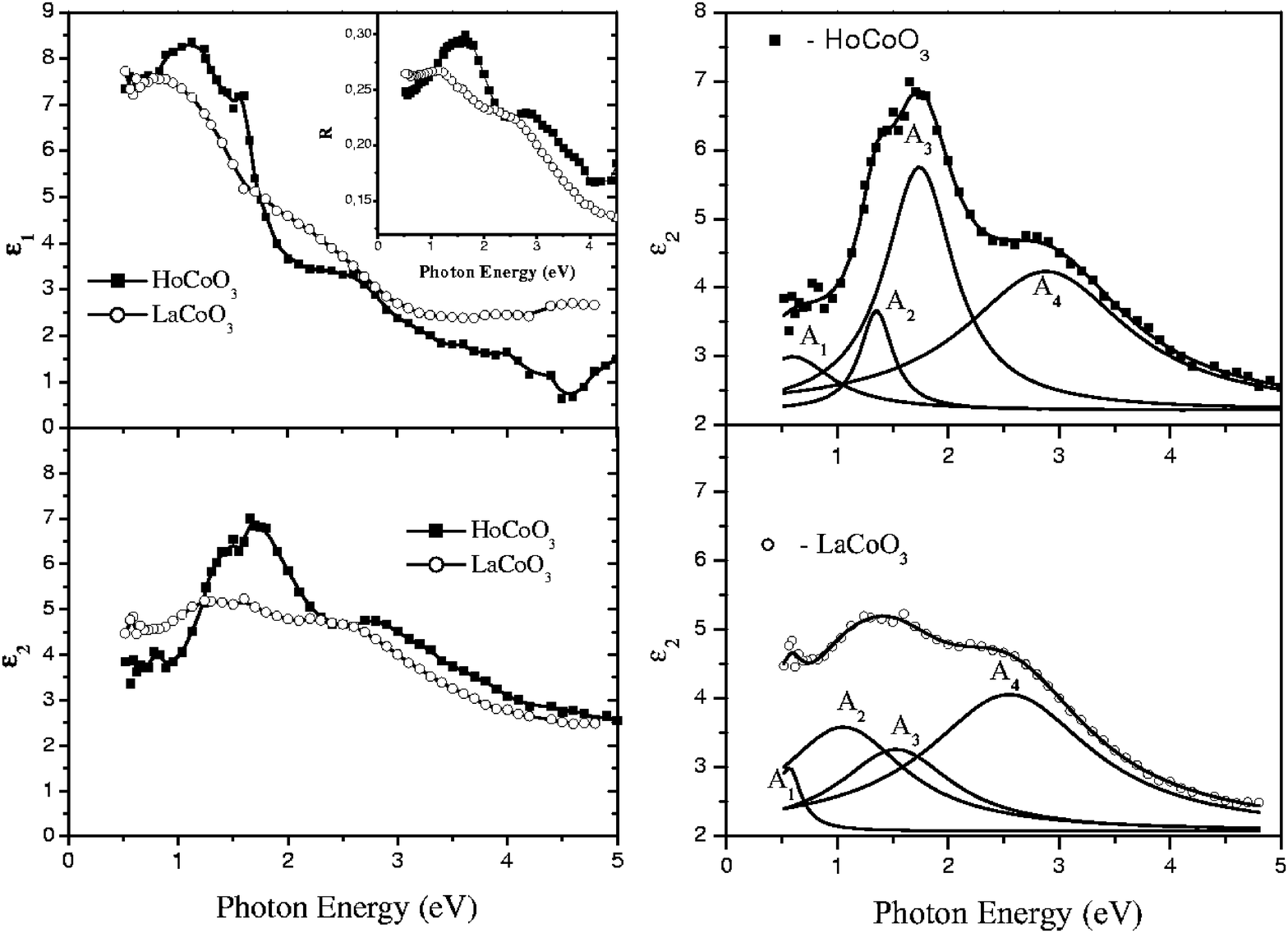}
\caption{\label{epsilon}
Left panel: the real $\varepsilon_1(\omega)$ and  imaginary $\varepsilon_2(\omega)$
parts of the dielectric function for LaCoO$_3$ and HoCoO$_3$. The inset in left panel
shows the reflectance $R(\omega)$.
Right panel: the variance analysis of $\varepsilon_2(\omega)$ for HoCoO$_3$
and LaCoO$_3$. }
\end{center}
\end{figure}

Changes in the optical charge gap values for the direct optical
transitions and consequently the shift of the main spectral
features with decrease of the rare earth ionic radii have been
already observed in the series of compounds RCoO$_3$ where R was
changed from La to Gd\cite{Yamaguchi-96}. It can be explained in a
following way. With decreasing of the rare earth ionic radii
Co-O-Co bond angles deviate much more from 180$^\circ$ and Co-O
bond length reduces. Since overlap between Co-3d and O-2p orbitals
is determined by Co-O distances and Co-O-Co bond angles, it
affects the $d-p-d$ interaction. As a result such distortions
cause the band width to decrease and hence band gap to increase.

Right panel of figure 1 shows a variance analysis of the imaginary part of the dielectric function
$\varepsilon_2(\omega)$, which has been expanded on 4 Lorentz oscillators
(A$_1$, A$_2$, A$_3$, A$_4$). Except the shift of oscillators maxima (on 0.04, 0.30, 0.21 and
0.33 eV respectively) to the high
energy region there is a strong enhancement (in 3 times) of
A$_3$ oscillators strength in HoCoO$_3$ in comparison
with LaCoO$_3$. At the same time other oscillator intensities are
changed not more than on 30$\%$.

Presumably, such strong enhancement of the optical absorption
can be associated with different spin-state
configurations of Co$^{3+}$ ions in these compounds. At room
temperature Co$^{3+}$ ions in HoCoO$_3$ are in LS, but
in LaCoO$_3$ a most part of them (about 80$\%$ from \cite{Tokura-98})
are in IS.  In this case absorption slackening in the range
1.3-2.3 eV in LaCoO$_3$ may be evidence of some transitions blocking
and as a result of suggestive changes in the band structure for IS
in comparison with LS. For LaCoO$_3$ similar effect of the spectral
feature intensity decreasing with temperature across spin-state transition
near 1 eV was observed in photoemission experiment\cite{Thomas-00}.

In order to compare experimental results with the band structure
calculations presented below, let us discuss the optical
conductivity spectra $\sigma(\omega)$ shown in figure 2. The optical
conductivity spectrum for our LaCoO$_3$ crystals is in a good
agreement with the literature data on numerical values and
dispersion\cite{Yamaguchi-96}. Optical properties for HoCoO$_3$
were investigated for the first time. The shift to the high energy
region and an appearance of addition absorption were observed in
optical conductivity $\sigma(\omega)$ as well as in
$\varepsilon_2(\omega)$ spectra. Optical properties of the series
of compounds RCoO$_3$ (R=La, Pr, Nd, Sm, Gd) were investigated by
Yamaguchi {\it et al} \cite{Yamaguchi-96} using reflectance
measurements with Kramers-Kronig analysis in the spectral range $E
< 2$ eV at T=9~K, where all compounds have the same LS state. Let
us analyze  this paper results and supplement them by our results
for HoCoO$_3$. If an intensity of the optical conductivity for
LaCoO$_3$ at 1.7 eV will be considered as 1, then for the series
of RCoO$_3$ compounds, where R=La, Pr, Nd, Sm, Gd, Ho it is
increased as a whole with a drastic drop on Nd like 1, 1.47, 1.30,
1.07, 1.27, 1.86. Similar tendency has the temperatures of the
spin-state transitions with decrease of the rare earth ionic
radii\cite{Liu-91}.

\begin{figure}
\begin{center}
 \epsfysize=100mm
 \epsfbox{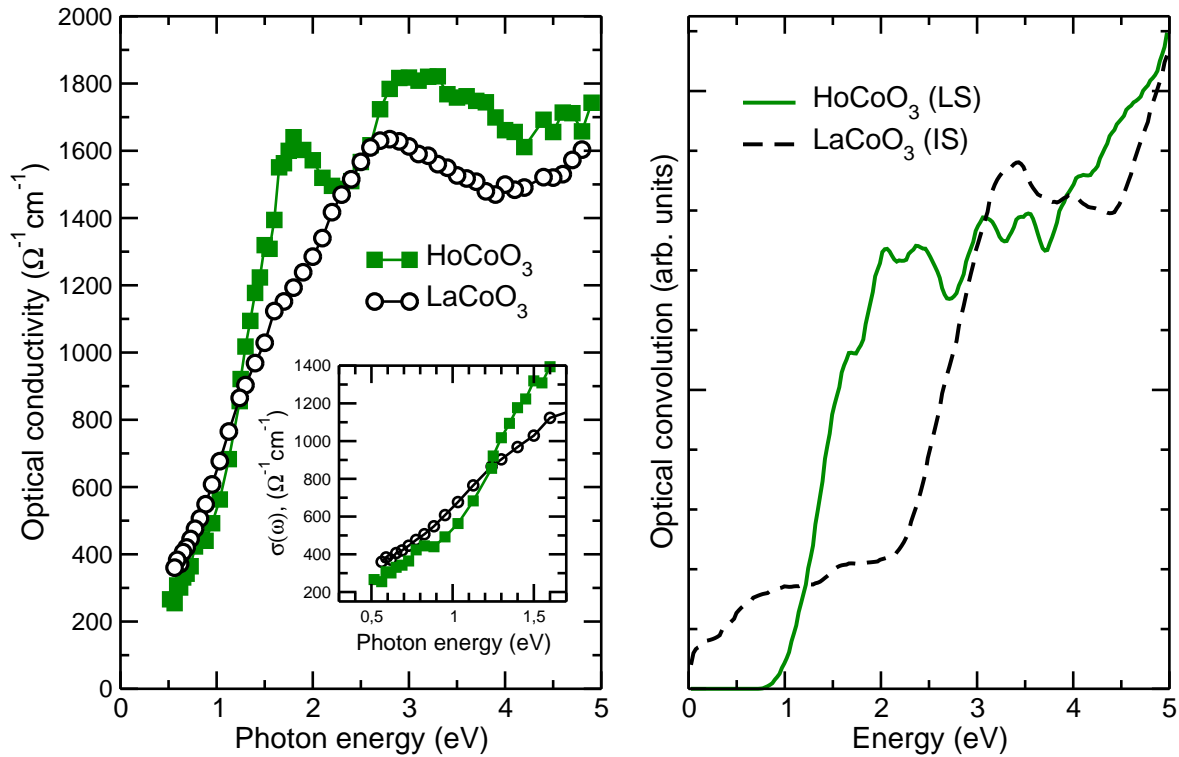}
 \caption {\label{3}Left panel: optical conductivity $\sigma(\omega)$ of HoCoO$_3$
and LaCoO$_3$ at 290~K. Right panel: calculated within the LDA+U
approach CPDOS for LaCoO$_3$ (dashed line) in
IS state and for HoCoO$_3$ (solid line) in LS state.
The Fermi level is zero energy.}
\end{center}
\end{figure}

Thus LS state becomes more stable with decrease
of the rare earth ionic radii and one can see, that the higher
temperature of transition from nonmagnetic LS state to
IS state is the bigger values of the spectral features
intensities are observed in the narrow spectral range
1.3 - 2.3 eV. That is naturally reflected in the electronic
structure reorganization: an additional channels of interband transitions
appearance.

\section{Computational details}
In order to find out an origin of the difference in the optical conductivity spectra
in LaCoO$_3$ and HoCoO$_3$ we've performed band structure
calculations within the LDA+U approximation\cite{Anisimov-91}. The calculation
scheme was realized in the framework of the linear muffin-tin orbitals
(LMTO)\cite{Andersen-75} method based on the Stuttgart TBLMTO-47
computer code.

Temperature was introduced in our calculations only by the change
of lattice parameters and atomic positions. Corresponding crystal
data for room temperature were taken from Radaelli {\it et
al.} (for LaCoO$_3$)\cite{Radaelli-02} and Liu {\it et al.} (for
HoCoO$_3$)\cite{Liu-91}. Co($4s$,$4p$,$3d$), O($3s$,$2p$) and
La,Ho($6s$,$6p$,$5d$) were included to the orbital basis set in
our calculations. In contrast to reference 10 almost empty La-$4f$
states as well as partially filled $4f$ states of Ho were treated
as pseudo-core states.

It is important to describe La-$4f$ states correctly because in the
present work we need to have as more precise density of states (DOS) as
possible, since it is used for the optical properties calculations.
An attempt to describe La-$4f$ states as valence one in the framework of LDA calculation
leads to the appropriate La-$4f$ states peaks appearance in the low energy
region near the Fermi level (see figures 3 - 5
in reference 10) in contrast to the experiment, where it placed at $\sim$ 9.0 eV
above E$_F$\cite{Chainani-92}.

However, we have found that the presence or the absence of La-$4f$
states in the basis set has not effect on total energies
difference of the various spin-state configurations. There are two
ways to describe La-$4f$ states properly: to apply LDA+U
correction on these states or to treat them as a pseudo-core.
We've chosen the second way.

On-site Coulomb interaction parameter and Stoner exchange parameter
were taken to be 7.0 eV and 0.99 eV respectively for Co-$3d$ shell.
The Brillouin-zone (BZ) integration in the course of the self-consistency
iterations was performed over a mesh of 27 {\bf k} points in the irreducible
part of the BZ. DOS as well as CPDOS (see below) were calculated by
the tetrahedron method with 512 {\bf k} points in the whole BZ.

There are two ways to calculate optical properties {\it ab-initio}.
First of all, one can calculate optical conductivity using the
Kubo formula\cite{Kubo-57,Wang-74} and matrix elements of momentum
operator $<{\bf k} n|-i\Delta_{\alpha}|{\bf k'}n>$.
But such rather difficult and time-consuming procedure
requires knowledge of the band structure with a better accuracy than that
provided by LMTO-ASA method\cite{Uspenski-83}.
Second, it is possible to calculate joint density of states (JDOS)\cite{Mazin-99}, defined as

\begin{equation}
J(h\omega)=\sum_{{\bf k}} N(\varepsilon_{\bf k})N(-(\varepsilon_{\bf k}+h\omega)),
\label{convolution}
\end{equation}

where $N(\varepsilon_{\bf k})$ is a total DOS. However, JDOS
calculations do not take into account selection rules and as a
result their comparison with an experiment usually is not good. In
this paper optical absorption is estimated to be proportional to
the convolution of partial density of states (CPDOS)

\begin{equation}
C(\omega)=\frac{1}{h\omega}
\sum_{\sigma}\int_{0}^{h\omega}N_{A}^{\sigma}(\varepsilon)N_{B}^{\sigma}(\varepsilon-h\omega)
d\varepsilon,
\label{convolution}
\end{equation}

where $N_{A}^{\sigma}(\varepsilon)$ and
$N_{B}^{\sigma}(\varepsilon-h\omega)$ are partial densities of
states (PDOS) respectively for the energy arguments above and
below of the Fermi level and $\sigma$ is spin projection. In this
way one should use a ''proper'' PDOS to satisfy the selection
rules. Due to such rules $N_{A}^{\sigma}(\varepsilon)$ and
$N_{B}^{\sigma}(\varepsilon-h\omega)$ should be the partial DOS of
the same atom, have the same spin index $\sigma$ and their orbital
quantum numbers should differ only on $\Delta l =$ $\pm$1. Thus,
this approximation can be compared with a conventional constant
matrix elements approximation\cite{Mazin-99}.

To this rule satisfy transitions $6s-6p$ and $6p-5d$ for Ho and La
ions, $3s-2p$ for O, $4s-4p$ and $4p-3d$ for Co ions.
On the contrary, the valence band in
both compounds is defined by O-$2p$ states which
hybridize with a partially filled Co-$3d$ band, but
O-$3s$ states are much far away from the Fermi level (approximately
on 30 eV higher) and hence oxygens $3s-2p$ transitions also should
be neglected.

Thus there are Co $4s-4p$, $4p-3d$ and La,Ho $6s-6p$, $6p-5d$ transitions which
can be taken into account in the calculation of (2). However,
there is a strong hybridization between Co-$4p$ and O-$2p$ states
due to a large spatial extension and a sizable overlap between
these orbitals on neighboring Co and O ions. So in this way
excitations from occupied Co-$4p$ into unoccupied Co-$3d$ band can
be considered in a certain way as O-$2p$ -- Co-$3d$ transitions.

\begin{figure}
\begin{center}
 \epsfysize=110mm
 \epsfbox{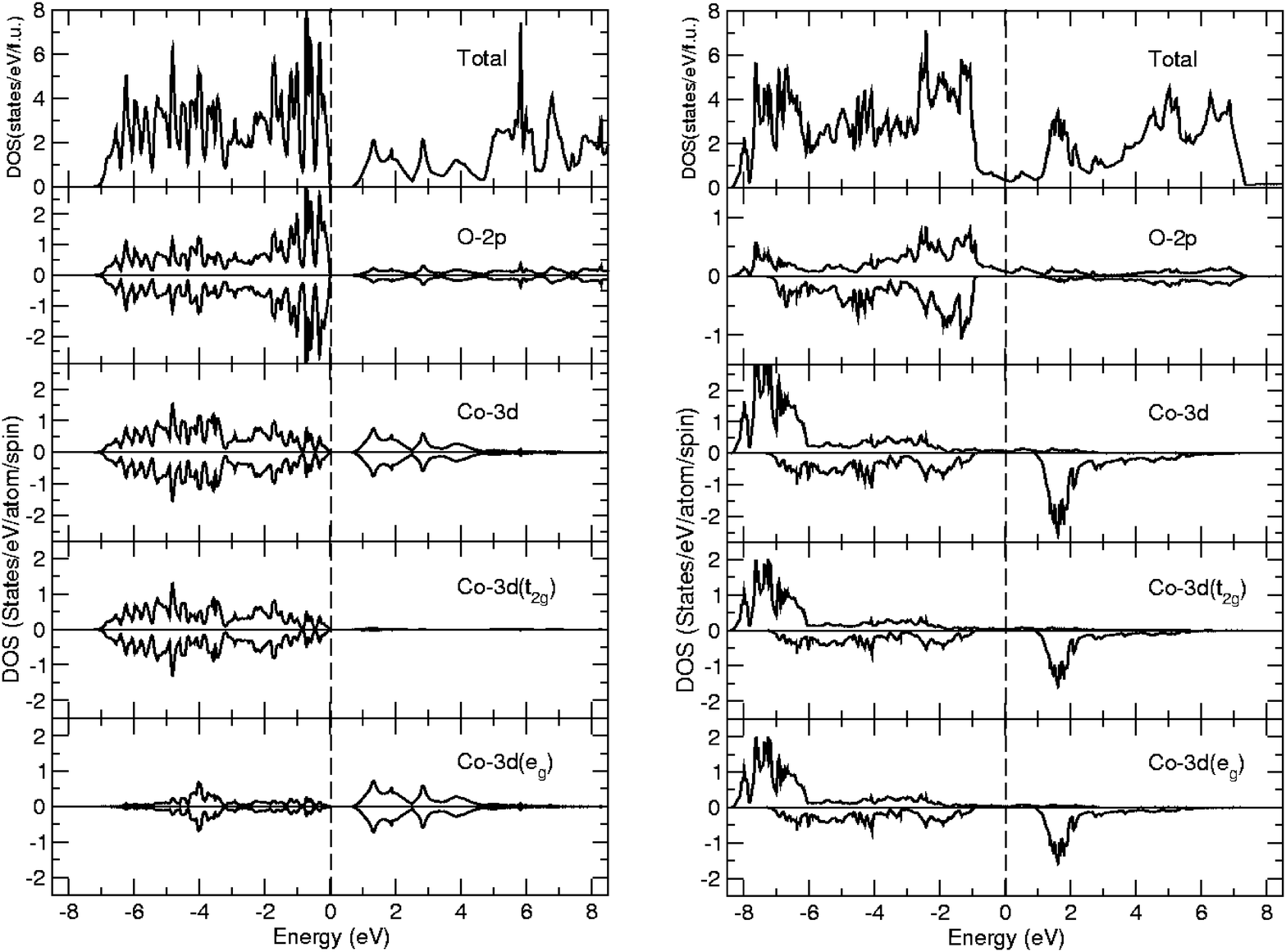}
 \caption {\label{ldaU} DOS calculated within the LDA+U approach for 
 HoCoO$_3$ LS state (t$^6_{2g}$e$^0_g$) of Co$^{3+}$ ions (left panel),
 LaCoO$_3$ IS state (t$^5_{2g}$e$^1_g$) of Co$^{3+}$ ions (right panel).
Parts of plots with positive (negative) ordinates
denote majority (minority) spin PDOS.
The Fermi level is zero energy.}
\end{center}
\end{figure}

\section{LDA+U results}
Recently, using the results of LDA+U calculations Nekrasov {\it et
al.}\cite{Nekrasov-03} have shown that a different spin-state
stabilizes in LaCoO$_3$ (IS) and in HoCoO$_3$ (LS) at room
temperature due to a chemical pressure happening with substitution
of La ions by small Ho ions and hence increasing of a crystal
field $t_{2g}-e_{g}$ energy splitting.

Calculated in the present work PDOS
 for IS state in LaCoO$_3$ and LS state in  HoCoO$_3$  are presented in
figure 3.
Both compounds should be insulators in the calculations at room
temperature. However, IS state for LaCoO$_3$ gave a metallic state in LDA+U
in contrast to the experiment\cite{Tokura-98}.
This contradiction have been explained
by a prediction of a possible orbital ordering of partially filled
e$_g$ orbitals of Co$^{3+}$ ions in IS state
\cite{Korotin-96}.
A band gap for HoCoO$_3$ in LS state is found to be 0.7 eV in
agreement with experiment, where it is estimated to be $\leq$ 0.7 eV (see left panel of
figure 2, inset).
Local magnetic moment on Co ion in LaCoO$_3$ is equal to 2.2 $\mu_{\rm B}$.

The results of optical convolution calculations are presented in
right panel of figure 2. Due to the presence of a sizable band gap in HoCoO$_3$
there are no electronic transitions below 0.7
eV, while LaCoO$_3$ is a metal in our calculation (insulator with
very small band gap $\sim$0.1 eV in
experiment\cite{Yamaguchi-96}). As a result in LaCoO$_3$ the electronic
excitations, which form optical conductivity, appear already at
very small energies. Drastic increase of the absorption is
observed at $\sim$ 2.3 eV for LaCoO$_3$ and at $\sim$ 1.1 eV for
HoCoO$_3$.

The origin of such different behavior one can find in
two interrelated features of O$-2p$ and Co$-3d$ PDOS for both compounds. 
From one side the $p-d$ hybridization for majority spin is
stronger for IS configuration because of partial filling of
$e_g-$band in IS in comparison with LS. It leads to the
increase of valence oxygen $2p$ band width in IS and to the decrease
of PDOS just below the Fermi level (compare left and right panels in figure 3). From other side
there is a sizable $\sim$ 2.3 eV gap in PDOS for Co$-3d$ and O$-2p$ minority spin,
because in general they are placed on different sides from the Fermi level 
and almost do not hybridize. 
Thus, there is a small intensity of the optical absorption in LaCoO$_3$ due to
only one (majority) spin contribution to this absorption 
in the low energy range till $\sim$ 2.3 eV.
In this case ''drastic growing'' of the intensity of  O$2p$ -- Co$3d$ transitions 
in optical conductivity spectra appears for IS state of Co$^{3+}$ ions in LaCoO$_3$  at
the higher energies than for LS Co$^{3+}$ in HoCoO$_3$.
Such shift and absorption intensity enhancement in the range 1.2-2.6 eV
results in the crossing of CPDOS for LaCoO$_3$ and HoCoO$_3$
at $\sim$ 1.2 eV in a good agreement with
experiment (see left panel of figure 2).

Thus, analysis of the LDA+U calculation results
shows that there are qualitative changes in the electronic structures
connected with the shift of the onset of the absorption and
absorption intensity enhancement due to the different spin-state
configuration at room temperature. It is important to note that
the structural modifications by itself without changing of the
spin-state configuration do not lead to the considerable changing
of the electronic structure (see figures 3 and 4 in reference 10).

\section{Conclusion}
In this paper, we've reported the results of the optical properties
investigations of isoelectronic compounds
LaCoO$_3$ and HoCoO$_3$  with the aim to clarify the influence
of the different spin-state stabilization
on electronic structure as a whole and on the optical
properties in particular.

We've measured the real $\varepsilon_1(\omega)$ and imaginary $\varepsilon_2(\omega)$
parts of the dielectric function, reflectance $R(\omega)$ and optical conductivity
and found several differences in the optical spectra between LaCoO$_3$ and HoCoO$_3$
at room temperature.

First of all, the shift of the main spectral features to the
high energy region on 0.3 eV associated with smaller rare earth
ionic radii and as a result with larger distortions in HoCoO$_3$ in
comparison with LaCoO$_3$ was found. As well, there is an increase of the
absorption in the range 1.3-2.3 eV in all kinds of spectra
in HoCoO$_3$, which can be attributed basing on the results of LDA+U calculations
to the different spin-state stabilization in these compounds.

We've performed convolution
of partial density of states computations obtained in the framework
of the LDA+U approach taking into account dipole-dipole
selection rules. As a result we've found qualitative changes
in the electronic structures, which is reflected in the optical
spectra as the shift of the onset of the absorption edge from
less than 0.1 eV in LaCoO$_3$ to 0.7 eV in HoCoO$_3$ and an absorption intensity
enhancement in a narrow spectral range 1.2-2.6 eV.
Such changes are assumed to be induced
by the different Co$^{3+}$ spin-state in these compounds at room
temperature.

\ack
The work was supported by the INTAS project No.01-0278
and the Russian Foundation for Basic Research through grants RFFI-01-02-17063
(VA, IN, MK) and RFFI-03-02-06026,
the grant of Ural Branch of the Russian Academy of Sciences for
Young Scientists, Grant of the President of Russia for
Young Scientists MK-95.2003.02 (IN).

\end{document}